\documentclass[conference,10pt]{IEEEtran}
\usepackage{epsfig,epsf,rotating,setspace,latexsym,amsmath,amssymb,amsfonts,bm,theorem,subfigure,epstopdf}
\usepackage{cite,authblk}
\usepackage{bbm}
\usepackage{algorithm}
\usepackage[noend]{algpseudocode}
\usepackage{color}
\usepackage{mathtools}
\usepackage{soul}

\algrenewcommand\algorithmicforall{\textbf{foreach}}
\algrenewcommand\algorithmicindent{.8em}

\newtheorem{lemma}{Lemma}

\newenvironment{Proof}[1]{\medskip\par\noindent{\bf Proof:\,}\,#1}{{\mbox{\,$\blacksquare$}\par}}

\IEEEoverridecommandlockouts
\allowdisplaybreaks

\begin{document}
 
\title{\LARGE \bf Timely Tracking of a Remote Dynamic Source Via \\ Multi-Hop Renewal Updates}
 
\author{Priyanka Kaswan \quad Sennur Ulukus\\
        \normalsize Department of Electrical and Computer Engineering\\
        \normalsize University of Maryland, College Park, MD 20742\\
        \normalsize  {\tt\small pkaswan@umd.edu} \quad {\tt\small ulukus@umd.edu}}
 
\maketitle

\begin{abstract}
We study the version age of information in a multi-hop multi-cast cache-enabled network, where updates at the source are marked with incrementing version numbers, and the inter-update times on the links are not necessarily exponentially distributed. We focus on the set of non-arithmetic distributions, which includes continuous probability distributions as a subset, with finite first and second moments for inter-update times. We first characterize the instantaneous version age of information at each node for an arbitrary network. We then explicate the recursive equations for instantaneous version age of information in multi-hop networks and employ semi-martingale representation of renewal processes to derive closed form expressions for the expected version age of information at an end user. We show that the expected age in a multi-hop network exhibits an additive structure. Further, we show that the expected age at each user is proportional to the variance of inter-update times at all links between a user and the source. Thus, end user nodes should request packet updates at constant intervals. 
\end{abstract}

\section{Introduction}\label{sect:introduction}

We consider a cache-enabled network consisting of a source node, server nodes and user nodes in a tree topology, with source as the root node and users as leaf nodes, as shown in Fig.~1. This type of topology is exhibited in multi-cast networks, where each server serves multiple base stations. The source gets updated according to an ordinary renewal process and uses a logical clock to mark the updates with an incrementing numeric value, which we refer to as the version number of the update packet. The user nodes attempt to retrieve the latest possible version update from the source through a sequence of cache-aided server nodes, such that updates on all links are forwarded according to ordinary renewal processes that are not necessarily Poisson processes. In this setting, version age of information is the natural choice of metric to quantify the freshness of information at the user nodes. At time $t$, if $W_i(t)$ is the latest version of a file available at node $i$ and $W_0(t)$ is the current version prevailing at the source, then the instantaneous version age at node $i$ at time $t$ is defined as $X_i(t)=W_0(t)-W_i(t)$. 

\begin{figure}[t]
\centerline{\includegraphics[width=0.6\linewidth]{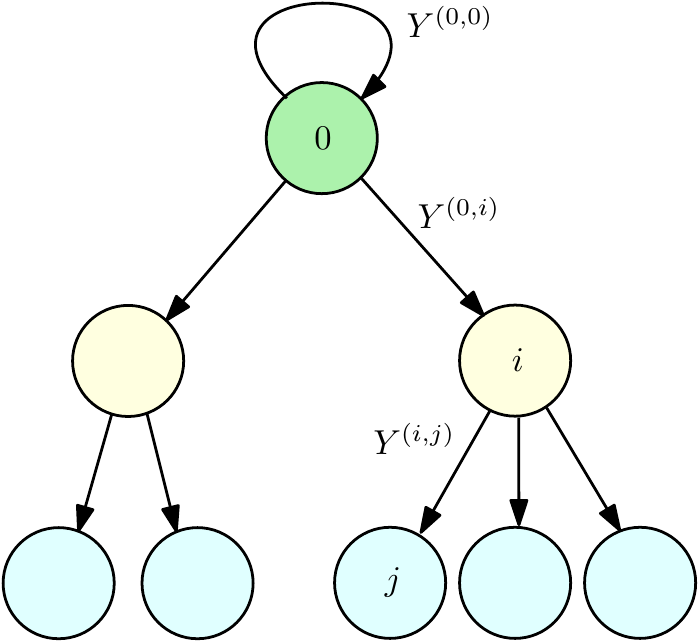}}
\caption{Multi-hop multi-cast tree network with version updates at source.}
\label{fig:tree networkV}
\vspace*{-0.4cm}
\end{figure}

Since the exponential distribution (or geometric distribution) is the only continuous (or discrete) probability distribution with memoryless property, most prior works studying timely information dissemination in networks heavily rely on these distributions \cite{yates18preempt, Kam18, Yates18parallel, talak17aoimultihop, selen13aoigossip, Yates17sqrt, bastopcu20_google, baturalp21clustergossip, bastopcu2020LineNetwork, kaswan_isit2021, Yates21gossip_traditional, Yates21gossip,maatouk22, kaswan22jammerring, kaswan22timestomping, mitra_allerton2022, Kaswan23reliable, delfani22_gossip_energy, elmagid23gossipagedist, Mitra23opportunisitic}. In this work, we focus on timely updating based on non-Poisson renewal processes in multi-hop networks and in this respect the related works are \cite{Kaswan23aoi_nonpoisson} and \cite{Yates20_moments}. For multi-hop networks operating under ordinary renewal processes, \cite{Kaswan23aoi_nonpoisson} derived analytical expression for traditional age of information while \cite{Yates20_moments} derived the distribution of traditional age of information for the case of strictly stationary age processes; see \cite[Section~V]{Yates20_moments}. 

There are significant differences in the analysis of version age of information (in this paper) from traditional age of information, by virtue of the additional renewal point process at the source superimposed on other renewal processes. This is because traditional age of information increases at unit rate, does not account for update process at the source, and consequently can be fully described at a node in terms of the time since the last update from the immediate server node and the age at the server node. On the other hand, version age of information is a discrete metric that gets incremented in steps of one whenever the source gets updated to a newer version, and involves counting of renewals at the source between update arrivals from the immediate server node. 

\begin{figure}[t]
\centerline{\includegraphics[width=0.95\linewidth]{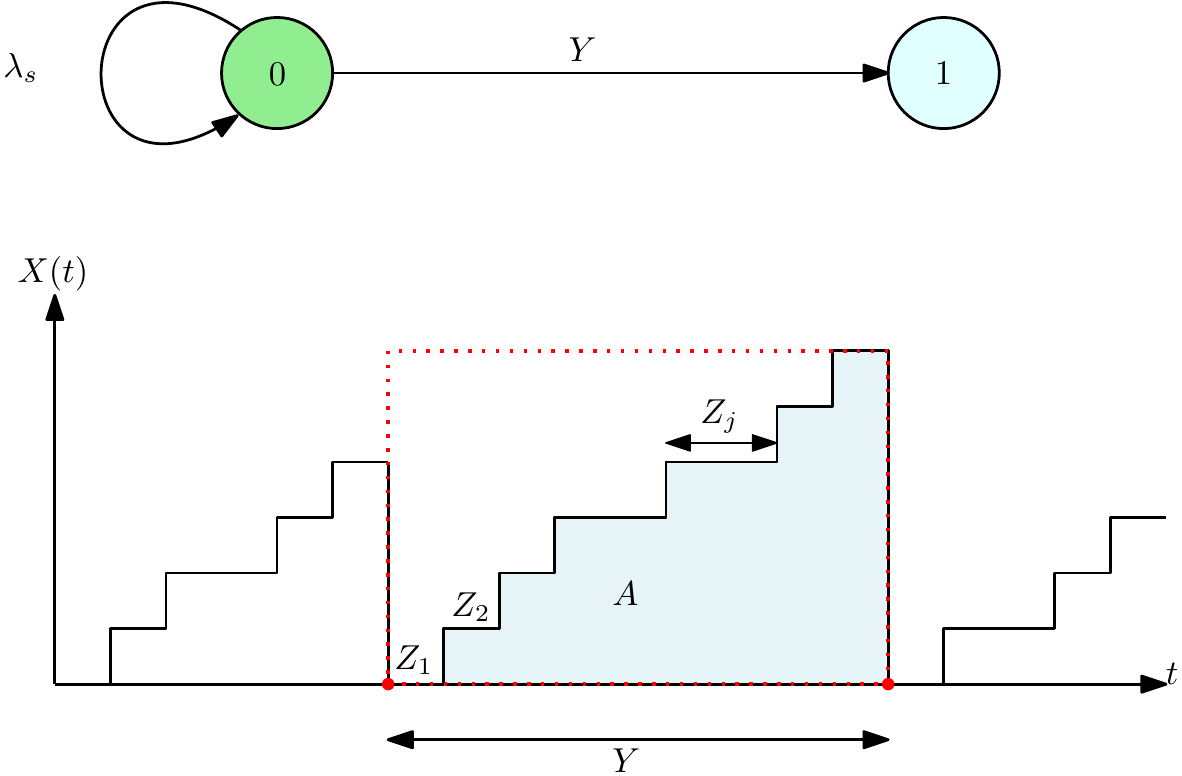}}
\caption{One-hop model, where the sources gets updates according to a Poisson process with rate $\lambda_s$ and updates arrive at node $1$ according to a renewal process with typical inter-renewal period $Y$.}
\label{fig:onehop_Poissonsource_V}
\vspace*{-0.4cm}
\end{figure}

Consider the single-hop network of Fig.~\ref{fig:onehop_Poissonsource_V}, where a user downloads packets from the source according to an ordinary renewal point process with inter-update times as positive i.i.d.~random variables, denoted by typical random variable $Y$ with non-arithmetic distribution $F$. In this respect, a distribution $F$ is called arithmetic (or periodic) if it is piecewise constant and its points of increase are contained in a set $\{0,d,2d, \ldots\}$ with the largest such $d>0$ being the span of such distribution. When $F$ is not arithmetic, it is called non-arithmetic, e.g., a distribution with a continuous part \cite{serfozo09}.

\begin{figure}[t]
\centerline{\includegraphics[width=0.95\linewidth]{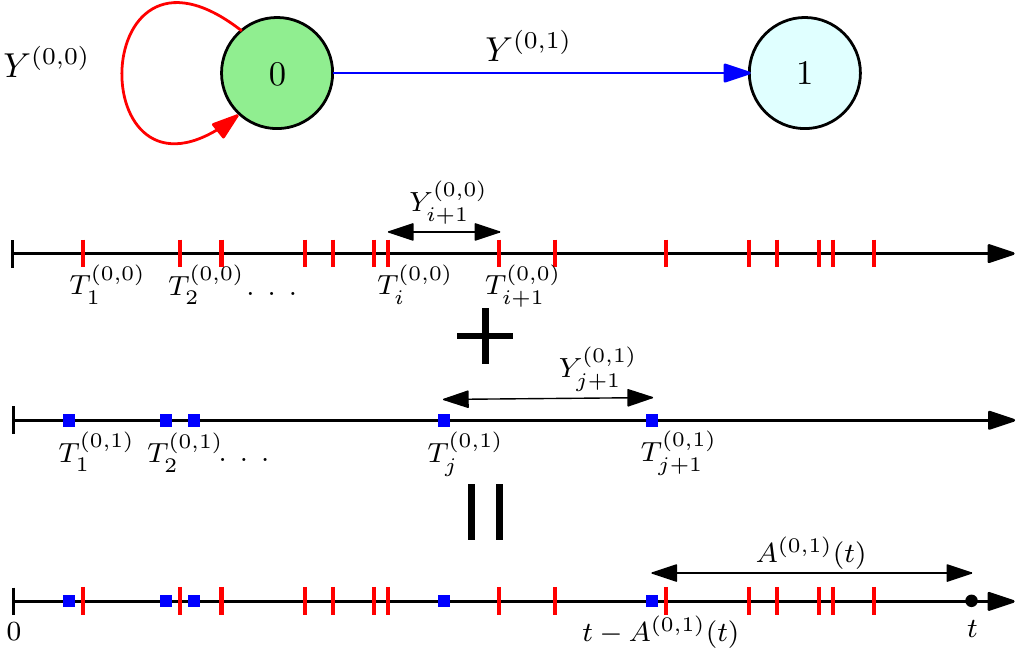}}
\caption{Superposition of renewal processes, $N^{(0,0)}(t)$ and $N^{(0,1)}(t)$, in the one-hop model.}
\label{fig:onehop_V}
\vspace*{-0.4cm}
\end{figure}

In Fig.~\ref{fig:onehop_Poissonsource_V}, first consider the simpler case when the source (node $0$) gets updated according to a Poisson process with rate $\lambda_s$. In this case, the instantaneous version age $X(t)$ at time $t$ at node $1$ is determined by the number of renewals at the source since node $1$ last downloaded a packet from the source. In Fig.~\ref{fig:onehop_Poissonsource_V}, $Y$ represents a typical inter-renewal interval between two consecutive downloads at node $1$, and $Z_j$ correspond to the inter-renewal interval $j$ between two updates at the source in interval $Y$, with $Z_1$ denoting the time between a download at node $1$ and the first update at the source since the download. Due to the memoryless property of Poisson process, $Z_1$ is exponentially distributed, like other $Z_j$s. Hence, $X(t)$ at any time $t$ within the particular inter-renewal interval $Y$ will only depend on the location of $t$ in $Y$ and random variables $Z_j$s local to $Y$, consequently, $X(t)$ qualifies as a renewal reward process. We define $A$ to be the accumulated reward in the inter-renewal interval $Y$, which corresponds to the area of the shaded region in Fig.~\ref{fig:onehop_Poissonsource_V}. We assume in this work that inter-renewal distributions of all renewal processes have finite first and second moments. Therefore, $\mathbb{E}[Y]<\infty$ and $\mathbb{E}[Y^2]<\infty$, and with probability 1, we have from \cite{gallager11}
\begin{align}\label{eqn:onehop_limit_expecV}
    \lim_{t \to \infty}\mathbb{E}[X(t)]=\frac{\mathbb{E}[A]}{\mathbb{E}[Y]}= \frac{\lambda_s\mathbb{E}[Y^2]}{2\mathbb{E}[Y]}
\end{align}
where the shaded area $A$ in Fig.~\ref{fig:onehop_Poissonsource_V} can be computed as in \cite{bastopcu20_google}. However, when the source gets updated according to a general renewal process, as in Fig.~\ref{fig:onehop_V}, in the absence of memoryless property of inter-update times, the distribution of $Z_1$ depends on the last source update instant in the previous inter-renewal interval, which prevents us from characterizing version age in the interval $Y$ independently of the past. 

\begin{figure}[t]
\centerline{\includegraphics[width=0.95\linewidth]{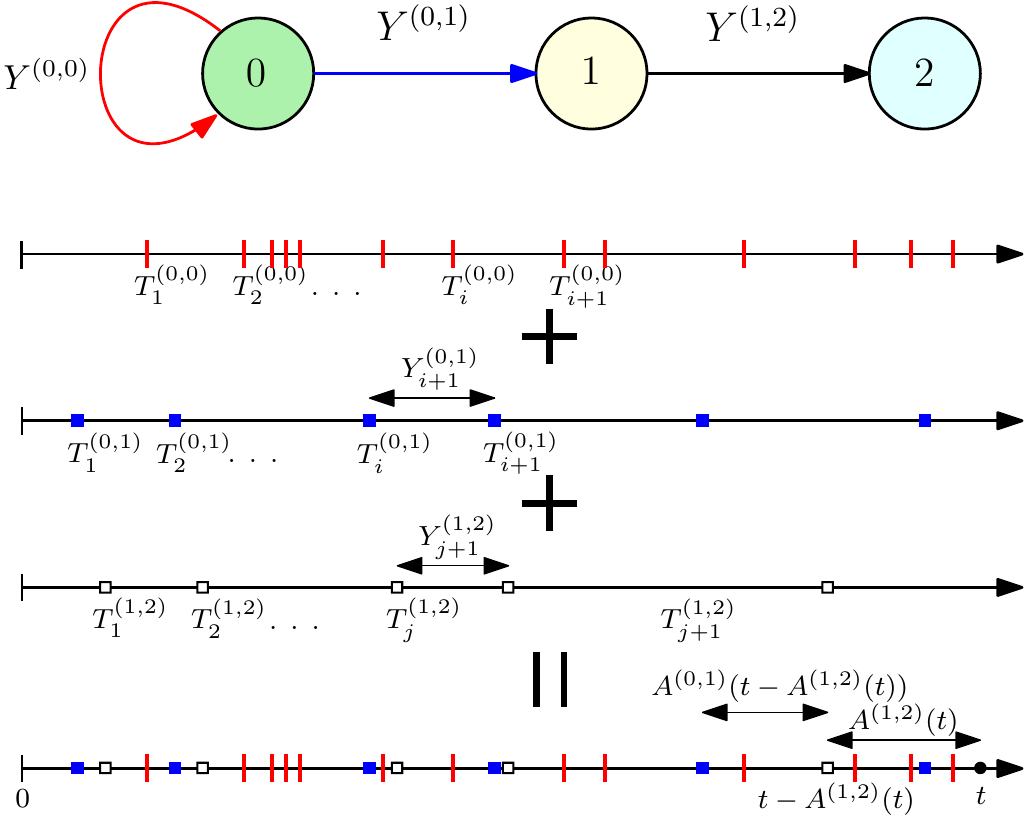}}
\caption{Superposition of source renewal process $N^{(0,0)}(t)$ on the two renewal processes, $N^{(0,1)}(t)$ and $N^{(0,1)}(t)$ in the two-hop model.}
\label{fig:twohopV}
\vspace*{-0.4cm}
\end{figure}

Further, if we added a second node to this model, this results in the two-hop model of Fig.~\ref{fig:twohopV}, where updates arrive at node $k$ from node $k-1$ at times $T^{(k-1,k)}_i$ according to renewal process $k$. If we consider a typical inter-renewal interval $[T^{(0,1)}_i,T^{(0,1)}_{i+1}]$, then in the absence of memoryless inter-update times, the distribution of packet arrival instant at node $2$ in this interval is dependent on when the last packet arrived at node $2$ in previous inter-renewal intervals, which prevents us from characterizing the age process in this interval independently of the past. Similarly, if we were to consider an inter-renewal interval $[T^{(1,2)}_j,T^{(1,2)}_{j+1}]$, then the user age at the beginning of this interval depends on the last packet arrival instant at node $1$ in prior intervals. Therefore, the age evolutions in different time intervals are correlated throughout the timeline, and an interesting question to ask here is, whether it is possible to somehow characterize the ensemble average of age at the end user in the regime of large $t$.

In this paper, we first attempt to characterize version age of information in cache-updating systems for ordinary renewal processes. Though getting expressions for the long-term expected version age proves difficult for general networks, we provide a closed form expression for the expected version age in multi-cast networks which exhibit a tree topology as shown in Fig.~\ref{fig:tree networkV}. To do so, we employ a pre-limit refinement of Blackwell's theorem \cite{daley19_martingale} using a semi-martingale representation of a renewal process $N(t)$, because the classic Blackwell's result, $\lim_{t \to \infty}\mathbb{E}[N(t+a)-N(t)]=\frac{a}{\mathbb{E}[Y]}$, only provides expected number of renewals for process $N(t)$ in a constant time interval $a$, which proves insufficient for version age analysis in this work. 

We show that the expected version age in a multi-hop network exhibits an additive structure. Further, we show that the expected version age at each user is proportional to the variance-to-mean ratio of the inter-update times at all links between a user and the source, and is inversely proportional to the mean of the inter-update times of renewal update process at the source. This implies that for a given average update rate, end users should request packet updates at constant intervals from their immediate servers to minimize their long-term expected version age of information, independent of the dynamics of the network. 

\section{Model and Notations}\label{sect:model_notations}

The source receives version updates according to a renewal process $N^{(0,0)}(t)$ and packets from node $i$ arrive at node $j$ on link $(i,j)$ according to a renewal process $N^{(i,j)}(t)$ . The corresponding finite random times $0\leq T^{(i,j)}_1 \leq T^{(i,j)}_2 \leq \ldots$ denote the  renewal times, such that the inter-arrival times $Y^{(i,j)}_n=T^{(i,j)}_n- T^{(i,j)}_{n-1}$ are positive i.i.d.~random variables with common distribution $F^{i,j}$, which is assumed to be non-arithmetic with finite first and second moments. Given $N^{(i,j)}(t)=\max\{n:T^{(i,j)}_n\leq t\}$, the regenerative process $A^{(i,j)}(t)=t-T^{(i,j)}_{N^{(i,j)}(t)}$ denotes the corresponding backward recurrence time (or age of renewal process) at time $t$, which is the time since the last renewal prior to $t$. Likewise, the regenerative process $B^{(i,j)}(t)=T^{(i,j)}_{N^{(i,j)}(t)+1}-t$ denotes the corresponding forward recurrence time (or residual renewal time of renewal process) at time $t$, which is the time to the next renewal after $t$. Note that $0 \leq A^{(i,j)}(t) \leq t$ (to be repeatedly used later). For more details, please see references \cite{serfozo09, gallager11}.

Consider a typical node $j$ in an arbitrary network of nodes and let $S_j$ denote the set of nodes from which packets arrive at node $j$. The most recent packet from node $i \in  S_j$ arrives at node $j$ before time $t$ at time instant $t-A^{(i,j)}(t)=T^{(i,j)}_{N^{(i,j)}(t)}$, at which point, node $j$ compares the version number of the arriving packet with the packet present at its cache, and discards the staler version in favor of the fresher version. 

Let $X_j(t)$ denote the instantaneous version age of information at node $j$ at time $t$. Then, $X_j(t)$ can be written as
\begin{align}\label{eqn:instantaneous_age_general_networksV}
    X_j(t)=\sum_{i\in S_j} \Big[\prod_{k\in S_j\backslash\{i\}}&\chi_{\{A^{(i,j)}(t)< A^{(k,j)}(t)\}}\Big] \nonumber \\
    \times\Big[\min&\big\{X_i(t-A^{(i,j)}(t)),X_j(t-A^{(i,j)}(t)\big\} \nonumber \\
    +N&^{(0,0)}(t) - N^{(0,0)}(t-A^{(i,j)}(t))\Big]
\end{align}
where $\chi_{\mathcal{A}}$ represents the indicator random variable for the measurable set $\mathcal{A}$ and $X_j(0)=0$. Since the source always has the latest packet, $X_0(t)=0$ at all times.\footnote{(\ref{eqn:instantaneous_age_general_networksV}) holds true for the case where the distributions of inter-update times do not have atom points. When the distributions have atom points, packets from different nodes might arrive at node $j$ at the same time with non-zero probability. This situation can be remedied by choosing a priority order for different incoming links, which would change some of the ``$<$'' to ``$\leq$'' in the indicator variable in (\ref{eqn:instantaneous_age_general_networksV}).}

In (\ref{eqn:instantaneous_age_general_networksV}), $\prod_{k\in S_j\backslash\{i\}}\chi_{\{A^{(i,j)}(t)< A^{(k,j)}(t)\}}$ corresponds to the scenario when the last packet that arrived at node $j$ before time $t$ came from node $i$, which would be the case when the backward recurrence times of all other relevant renewal processes at time $t$ are larger than $A^{(i,j)}(t)$. The last term $N^{(0,0)}(t) - N^{(0,0)}(t-A^{(i,j)}(t))$ in (\ref{eqn:instantaneous_age_general_networksV}) comes from the fact that version age at node $j$ increments by one every time the source gets updated post the last packet arrival at $t-A^{(i,j)}$.

In the next step, $\min\{X_i(t-A^{(i,j)}(t)),X_j(t-A^{(i,j)}(t)\}$, which is a minimum over two age processes, can be further characterized in a manner similar to (\ref{eqn:instantaneous_age_general_networksV}) by accounting for arrivals at nodes $\{i,j\}$ from the set $S_i\cup S_j$, and the corresponding expression will have terms of the form that involve taking a minimum over three age processes, for example $\min\{X_i(t'),X_j(t'),X_k(t')\}$ with $t'=t-A^{(i,j)}(t)- A^{(k,j)}(t-A^{(i,j)}(t))$. By recursively repeating this process we finally encounter the expression $\min\{X_1(t''),X_2(t''),\ldots,X_n(t'')\}$, $t''=t-\Delta(t)$, where $\Delta(t)$ represents a stochastic process whose exact expression depends on the network topology. Since the source node is the only node external to the set of $n$ nodes, the last $\min$ expression can be completely defined in terms of the backward recurrence times of the form $A^{(0,\ell)}(t''')$ for all $\ell$ with $0\in S_{\ell}$. This recursive approach will become more clear for multi-cast networks in Section~\ref{sect:age_treenetworks}.

On first glance, this might give an impression that since all renewal processes and their associated recurrence times are independent processes, by reducing $X_j(t)$ to a function composed purely of backward recurrence times, one could conveniently compute the expectation of $X_j(t)$. However, note that, in the first and second steps of the recursion above, we encountered the term $A^{(k,j)}(t)$ in the product of indicator variables in (\ref{eqn:instantaneous_age_general_networksV}), and the term $A^{(k,j)}(t-A^{(i,j)}(t))$ in the definition of $t'$. Though both terms correspond to the same renewal process $N^{(k,j)}(t)$, these backward recurrence times could be correlated through time which complicates analysis. 

However, this complication does not arise if we assume that each node in the network has only one incoming link, as shown in the tree network of Fig.~\ref{fig:tree networkV}. This is because, $|S_j|=1$, and the product term in (\ref{eqn:instantaneous_age_general_networksV}) vanishes. Additionally, since packets now arrive at node $j$ from a single preceding node $i$, $X_i(t)\leq X_j(t)$ for all $t$, this simplifies the $\min$ term as follows
\begin{align}
    \min \big \{ X_i(  t \! - \! A^{(i,j)}(t) ),X_j( t \! - \! A^{(i,j) \!}(t) )\big\} = X_i( t \! - \! A^{(i,j)}(t) )
\end{align}

In the next section, we derive a closed form expression for the long-term expected version age $\lim_{t \to \infty}\mathbb{E}[X_j(t)]$ at each node $j$ in networks that have a tree topology.

\section{Age in Networks with Tree Structure}\label{sect:age_treenetworks}

\subsection{One-Hop Network}

We consider a single-hop network, where the source gets updated according to a renewal process that is not necessarily Poisson; see Fig.~\ref{fig:onehop_V}.
At time $t$, the last packet arrival at node $1$ from node $0$ happens at time $t-A^{(0,1)}(t)$, therefore, the instantaneous version age at node $1$ depends on the number of version updates at the source in the interval $(t-A^{(0,1)}(t),t]$, thus giving
\begin{align} \label{eqn:X1(t)_onehop_V_instant}
    X_1(t) =  N^{(0,0)}(t) - N^{(0,0)} (t \! - \!A^{(0,1)} (t)) + X_0(t \! - \! A^{(0,1)} (t))
\end{align}

\begin{lemma}\label{lemma:composition_2stoch_procs}
    Given independent stochastic processes $S_1(t)$ and $S_2(t)$ with $\sup_{t\geq 0}\left|\mathbb{E}[S_1(t)]\right|<\infty$ and $0\leq S_2(t)\leq t$, such that, $\lim_{t \to \infty}\mathbb{E}[S_1(t)]$ and $\lim_{t \to \infty}\mathbb{E}[S_2(t)]$ exist, we have
    \begin{align}
        \lim_{t \to \infty}\mathbb{E}[S_1(t-S_2(t))]=\lim_{t \to \infty}\mathbb{E}[S_1(t)]
    \end{align}
\end{lemma}
Lemma~\ref{lemma:composition_2stoch_procs} is first presented and proved in \cite{Kaswan23aoi_nonpoisson}.

\begin{lemma}\label{lemma:N(t)_semimartingale}
    Let $N(t)$ be a renewal process with i.i.d.~inter-renewal times, denoted by typical random variable $Y$, with non-arithmetic distribution and finite first and second moments. Let $S(t)$ be a stochastic process that is independent of $N(t)$, such that, $0\leq S(t) \leq t$ and $\lim_{t \to \infty}\mathbb{E}[S(t)]$ exists. Then,
    \begin{align}
        \lim_{t \to \infty}\mathbb{E}[N(t)-N(t-S(t))]=\frac{\lim_{t \to \infty}\mathbb{E}[S(t)]}{\mathbb{E}[Y]}
    \end{align}
\end{lemma}

\begin{Proof}
Taking $\mu=\mathbb{E}[Y]$, \cite{daley19_martingale} provides the following semi-martingale representation for a renewal process $N(t)$,
\begin{align}
    N(t)=\frac{t+B(t)}{\mu}+M(t)
\end{align}
where $B(t)$ is the forward recurrence time associated with the renewal process $N(t)$ and $M(t)=N(t)+1-\frac{T_{N(t)+1}}{\mu}$ is a martingale. Let $m(t)=\mathbb{E}[N(t)]$ and $b(t)=\mathbb{E}[B(t)]$, then since $\mathbb{E}[Y]<\infty$ and $\mathbb{E}[Y^2]<\infty$, renewal reward theorem \cite{gallager11} gives $\lim_{t\to \infty}b(t)=\frac{\mathbb{E}[Y^2]}{2\mathbb{E}[Y]}=b_1$. This implies that there exists $T$, such that, for all $t \geq T$, $b(t)<b_1 + \epsilon$, for some $\epsilon >0$. Further, for any $t < T$, $B(t) \leq (T-t)+B(T) \leq T+b_1 + \epsilon$, in the worst case, no renewal occurs in the time interval $(t,T)$ which leads to $B(t) =(T-t)+B(T)$. Hence,
\begin{align}
    \sup_{t \geq 0}|b(t)| \leq T+b_1+\epsilon < \infty
\end{align}

Using the fact that $\mathbb{E}[M(t)]=\mathbb{E}[M(0)]=0$ for all $t$ (see Wald identity, \cite{gallager11}) since $M(t)$ is a martingale, we get
\begin{align}
    \mathbb{E}[N(t)&-N(t-S(t))] \nonumber \\
    &=\mathbb{E}\left[\frac{t+B(t)}{\mu} - \frac{t-S(t)+B(t-S(t))}{\mu}\right] \\
    &= \mathbb{E}\left[\frac{S(t)+B(t)-B(t-S(t))}{\mu}\right]
\end{align}
Taking limit $t \to \infty$ on both sides and using Lemma 1 with $S_1(t)=B(t)$ and $S_2(t)=S(t)$ gives the desired result.
\end{Proof}

Coming back to computing $\lim_{\to \infty}\mathbb{E}[X_1(t)]$ in Fig.~\ref{fig:onehop_V}, using (\ref{eqn:X1(t)_onehop_V_instant}) and Lemma~\ref{lemma:N(t)_semimartingale} along with $X_0(t-A^{(0,1)}(t))=0$, we have 
\begin{align}
    \lim_{t \to \infty}\mathbb{E}[X_1(t)] & = \frac{\lim_{t \to \infty}\mathbb{E}[ A^{(0,1)}(t)]}{\mathbb{E}[Y^{(0,0)}]}  \\
    & = \frac{\mathbb{E}[ \left( Y^{(0,1)} \right)^2]}{2\mathbb{E}[Y^{(0,0)}]\mathbb{E}[Y^{(0,1)}]}
\end{align}
Note that if both the processes in Fig.~\ref{fig:onehop_V} are Poisson, then the expected version age at the user node is known to be $\frac{\lambda_s}{\lambda}$\cite{Yates21gossip,bastopcu20_google}. Therefore, it is interesting to note that $\lambda_s$ here is the proxy for $\frac{1}{\mathbb{E}[Y^{(0,0)}]}$, while $\lambda$ is the proxy for $\frac{2\mathbb{E}[Y^{(0,1)}]}{\mathbb{E}[ \left( Y^{(0,1)} \right)^2]}$.

\subsection{Two-Hop Network}

Consider the two-hop network in Fig.~\ref{fig:twohopV} where we wish to determine the long-term expected age at node $2$. Then, the instantaneous version age $X_2(t)$ can be written as
\begin{align}\label{eqn:X_2_two_hop_recursiveV}
    X_2(t)=&  N^{(0,0)}(t) - N^{(0,0)}(t-A^{(1,2)}(t))  \nonumber\\
    &+X_1(t-A^{(1,2)}(t))      
\end{align}
where $X_1(t-A^{(1,2)}(t))$ in turn can be expressed as
\begin{align}\label{eqn:X_1_two_hop_recursiveV}
    X_1(t-&A^{(1,2)}(t)) \nonumber\\
    =&N^{(0,0)}(t-A^{(1,2)}(t)) \nonumber\\
    &- N^{(0,0)}(t-A^{(1,2)}(t)-A^{(0,1)}(t-A^{(1,2)}(t))) \nonumber\\ 
    &+X_0(t-A^{(1,2)}(t)-A^{(0,1)}(t-A^{(1,2)}(t)))    
\end{align}
Let us define 
\begin{align}
    \Delta_1(t)&=A^{(1,2)}(t)\label{eqn:delta1_2hopV}\\
    \Delta_2(t)&=A^{(0,1)}(t-A^{(1,2)}(t))\label{eqn:delta2_2hopV}
\end{align}
Substituting (\ref{eqn:X_1_two_hop_recursiveV}), (\ref{eqn:delta1_2hopV}) and (\ref{eqn:delta2_2hopV}) in (\ref{eqn:X_2_two_hop_recursiveV}) and using $X_0(t-\Delta_1(t)-\Delta_2(t))=0$, we get
\begin{align}
    X_2(t)=& N^{(0,0)}(t) - N^{(0,0)}(t-\Delta_1(t)) + N^{(0,0)}(t-\Delta_1(t))\nonumber\\
    &- N^{(0,0)}(t-\Delta_1(t)-\Delta_2(t)) \\
    =&N^{(0,0)}(t) - N^{(0,0)}(t-\Delta_1(t)-\Delta_2(t)) \label{eqn:twohop_instantageV}
\end{align}

To compute the expectation in (\ref{eqn:twohop_instantageV}), $\lim_{t \to \infty}\mathbb{E}[X_2(t)]$, we use Lemma~\ref{lemma:N(t)_semimartingale}, which requires computing the terms $\mathbb{E}[\Delta_1(t)]$ and $\mathbb{E}[\Delta_2(t)]$ at $t \to \infty$. The backward recurrence time $A^{(1,2)}(t)$ has the following limiting expectation \cite{gallager11}
\begin{align}\label{eqn:2hop_limit_delta1V}
    \lim_{t \to \infty}\mathbb{E}[\Delta_1(t)]=\lim_{t \to \infty}\mathbb{E}[A^{(1,2)}(t)]=\frac{\mathbb{E}\left[\left(Y^{(1,2)}\right)^2\right]}{2\mathbb{E}\left[Y^{(1,2)}\right]}
\end{align}
Likewise, we have
\begin{align}
    \lim_{t \to \infty}\mathbb{E}[A^{(0,1)}(t)]=\frac{\mathbb{E}\left[\left(Y^{(0,1)}\right)^2\right]}{2\mathbb{E}\left[Y^{(0,1)}\right]}
\end{align}
Since the limit $\lim_{t \to \infty}\mathbb{E}[A^{(0,1)}(t)]$ exists, there exists $T$ such that for all $t>T$, $\mathbb{E}[A^{(0,1)}(t)]<\lim_{t \to \infty}\mathbb{E}[A^{(0,1)}(t)]+\epsilon$ for some $\epsilon>0$. Further, since $0\leq A^{(0,1)}(t) \leq t$ by definition, we have $\mathbb{E}[A^{(0,1)}(t)]<T$ for $t\leq T$. Hence, 
\begin{align}\label{eqn:2hop_supremum_01proofV}
    \sup_{t\geq 0}\left|\mathbb{E}[A^{(0,1)}(t)]\right|\leq \max\big\{\lim_{t \to \infty}\mathbb{E}[A^{(0,1)}(t)]+\epsilon,T\big\}<\infty
\end{align}
Hence, by Lemma~\ref{lemma:composition_2stoch_procs},
\begin{align}\label{eqn:2hop_limit_delta2V}
    \lim_{t \to \infty}\mathbb{E}[\Delta_2(t)]=\lim_{t \to \infty}\mathbb{E}[A^{(0,1)}(t)]=\frac{\mathbb{E}\left[\left(Y^{(0,1)}\right)^2\right]}{2\mathbb{E}\left[Y^{(0,1)}\right]}
\end{align}
Since $0\leq \Delta_1(t)+\Delta_2(t) \leq t$ and $\lim_{t \to \infty}\mathbb{E}[\Delta_1(t)+\Delta_2(t)]$ exists, by Lemma~\ref{lemma:N(t)_semimartingale} the long-term expected age at node $2$ is 
\begin{align}
    \lim_{t \to \infty}&\mathbb{E}[X_2(t)]=\frac{\lim_{t \to \infty}\mathbb{E}[\Delta_1(t)+\Delta_2(t)]}{\mathbb{E}[Y^{(0,0)}]}\\
    & =\frac{1}{\mathbb{E}[Y^{(0,0)}]}\left(\frac{\mathbb{E}\left[\left(Y^{(1,2)}\right)^2\right]}{2\mathbb{E}\left[Y^{(1,2)}\right]}+\frac{\mathbb{E}\left[\left(Y^{(0,1)}\right)^2\right]}{2\mathbb{E}\left[Y^{(0,1)}\right]} \right) 
\end{align}

Interestingly, the age at node $2$ is determined by the sum of independent contributions of links in the path from node $0$ to node $2$, divided by $\mathbb{E}[Y^{(0,0)}]$. In general, for tree networks, only the links involved in the path between the source and an end user are critical to the age dynamics of the end user apart from the update process at the source, and therefore, in the next subsection, we study $n$-hop linear networks.

\subsection{Multi-Hop Network}

\begin{figure}[t]
\centerline{\includegraphics[width=0.95\linewidth]{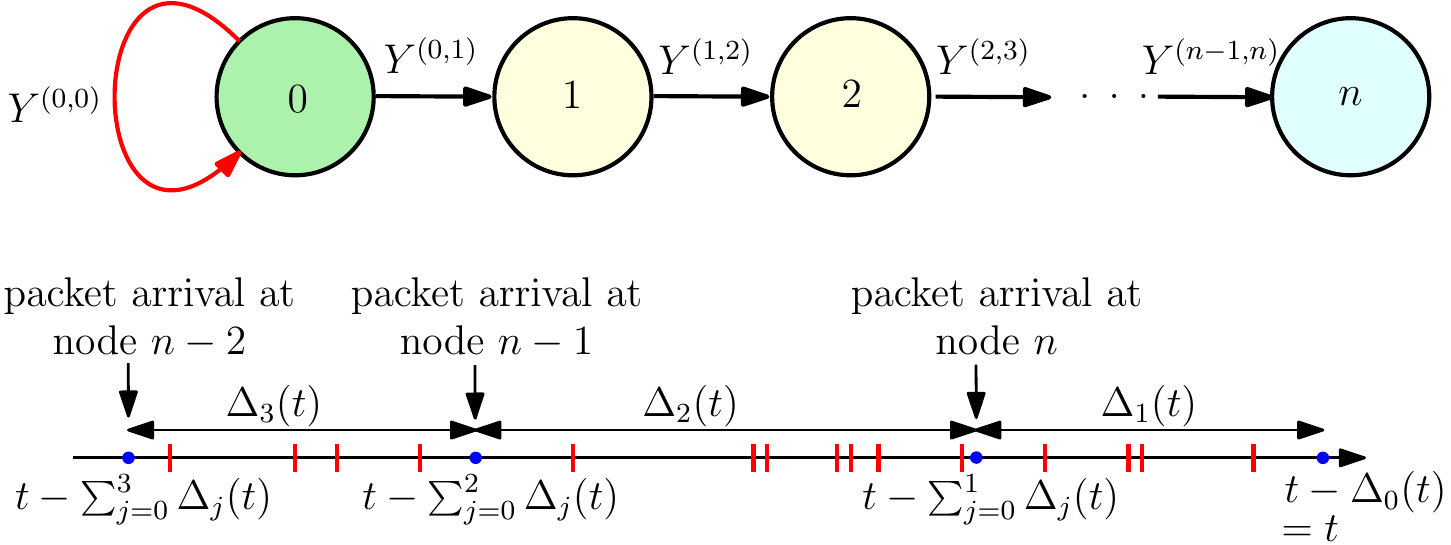}}
\caption{Time segments $\Delta_{i}(t)$ for $i\in \{ 1,\ldots,n\}$ in $n$-hop model. The points marked by red represent version updates at the source.}
\label{fig:n_hop_modelV}
\vspace*{-0.4cm}
\end{figure}

Consider the $n$-hop network of Fig.~\ref{fig:n_hop_modelV} where we wish to determine the long-term expected age at node $n$ of the network. We define time segments $\Delta_{i}(t)$, $i\geq 1$ through the following recurrence equation
\begin{align}\label{eqn:multi-hop_deltarecursieveV}
    \Delta_{i}(t)&=A^{(n-i,n-i+1)}(t-\sum_{j=0}^{i-1}\Delta_j(t))
\end{align}
with $\Delta_0(t)=0$, see Fig.~\ref{fig:n_hop_modelV}. Note that $\Delta_{i}(t)$ is smaller than $t-\sum_{j=0}^{i-1}\Delta_j(t)$ by definition of $A^{(n-i,n-i+1)}(t)$. Similar to (\ref{eqn:X_2_two_hop_recursiveV}), the instantaneous age $X_n(t)$ at node $n$ can be written as
\begin{align}\label{eqn:multi-hop_Xnt_rawV}
    X_n(t)=& N^{(0,0)}(t) - N^{(0,0)}(t-A^{(n-1,n)}(t)) \nonumber\\
    &+X_{n-1}(t-A^{(n-1,n)}(t))
\end{align}
This can be alternately represented as
\begin{align}
    X_n(t-\Delta_0(t))=& N^{(0,0)}(t) - N^{(0,0)}(t-\Delta_0(t)-\Delta_1(t)) \nonumber\\
    &+X_{n-1}(t-\Delta_0(t)-\Delta_1(t)) 
\end{align}

In the next step, $X_{n-1}(t-A^{(n-1,n)}(t))$ of (\ref{eqn:multi-hop_Xnt_rawV}) will be again characterized in a similar manner and the full set of equations encountered in this recursive approach is of the form
\begin{align}\label{eqn:Xnit_recursiveeqnV}
    X_{n-i}(t-\sum_{j=0}^{i}\Delta_j(t))=& N^{(0,0)}(t-\sum_{j=0}^{i}\Delta_j(t))\nonumber\\
    & - N^{(0,0)}(t-\sum_{j=0}^{i+1}\Delta_j(t)) \nonumber\\
    &+ X_{n-i-1}(t-\sum_{j=0}^{i+1}\Delta_j(t)) 
\end{align}
for $0\leq i\leq n-1$ with $X_0(t-\sum_{j=1}^{n}\Delta_j(t))=0$ as node $0$ represents the source node. Then, it follows from (\ref{eqn:Xnit_recursiveeqnV}) that
\begin{align}\label{eqn:nhop_Xnt_sum_instantaneousV}
    X_n(t)=N^{(0,0)}(t) - N^{(0,0)}(t-\sum_{j=1}^{n}\Delta_j(t))
\end{align}

Similar to (\ref{eqn:2hop_limit_delta1V}), we have 
\begin{align}
    \!\!\!\!\lim_{t \to \infty}\mathbb{E}[\Delta_1(t)]=\lim_{t \to \infty}\mathbb{E}[A^{(n-1,n)}(t)]=\frac{\mathbb{E}\left[\left(Y^{(n-1,n)}\right)^2\right]}{2\mathbb{E}\left[Y^{(n-1,n)}\right]}\!\!\!
\end{align}
Further, using the approach of (\ref{eqn:2hop_supremum_01proofV}), we get $\sup_{t\geq 0}\left|\mathbb{E}[A^{(n-i,n-i+1)}(t)]\right|<\infty$. Since $\sum_{j=0}^{i-1}\Delta_j(t)\leq t$, we can prove $\lim_{t \to \infty}\mathbb{E}[\Delta_i(t)]=\lim_{t \to \infty}\mathbb{E}[A^{(n-i,n-i+1)}(t)]$ recursively for $i=2,3,\ldots,n$ from (\ref{eqn:multi-hop_deltarecursieveV}) using Lemma~\ref{lemma:composition_2stoch_procs}.

Hence, from (\ref{eqn:nhop_Xnt_sum_instantaneousV}), we obtain
\begin{align}
    \lim_{t \to \infty}\mathbb{E}[X_n(t)]=&\frac{\sum_{j=1}^{n}\lim_{t \to \infty}\mathbb{E}[\Delta_j(t)] }{\mathbb{E}[Y^{(0,0)}]}\\
    =&\frac{1}{\mathbb{E}[Y^{(0,0)}]}\sum_{j=1}^{n}\frac{\mathbb{E}\left[\left(Y^{(n-j,n-j+1)}\right)^2\right]}{2\mathbb{E}\left[Y^{(n-j,n-j+1)}\right]}\label{eqn:nhop_limit_expec_age_node_n_V}
\end{align}

Interestingly, the age at node $n$ depends on independent contributions of the intermediate links $(i,i+1)$, $0\leq i \leq n-1$ and is invariant to the ordering of these links. Hence, each node can minimize its age by optimizing its individual packet request renewal process, irrespective of the statistical properties of other nodes and links in the network. Since the constant random variable has zero variance, for a fixed mean $\mathbb{E}[Y^{(n-j,n-j+1)}]$, (\ref{eqn:nhop_limit_expec_age_node_n_V}) hints that all nodes should request packets at near constant time intervals to reduce variance. Further, the age at node $n$ is inversely proportional to $\mathbb{E}[Y^{(0,0)}]$, implying the version age at nodes would be larger for a fast updating source on average. 

Further, if all renewal processes in Fig.~\ref{fig:n_hop_modelV} were Poisson, with $\lambda_s$ as the source update rate and $\lambda_j$ as the link $(n-j,n-j+1)$ update rate, respectively, then (\ref{eqn:nhop_limit_expec_age_node_n_V}) simplifies to
\begin{align}
    \lim_{t \to \infty}\mathbb{E}[X_n(t)]=\lambda_s\sum_{j=1}^{n}\frac{1}{\lambda_j}
\end{align}
which also results from \cite[Thm.~1]{Yates21gossip} or \cite[Eqn.~(11)]{baturalp21clustergossip} and has an interesting parallelism with \cite[Thm.~2]{yates18preempt}.

\section{Numerical Results}\label{sect:numerical_ResultsV}

\begin{figure}[t]
\centerline{\includegraphics[width=0.9\linewidth]{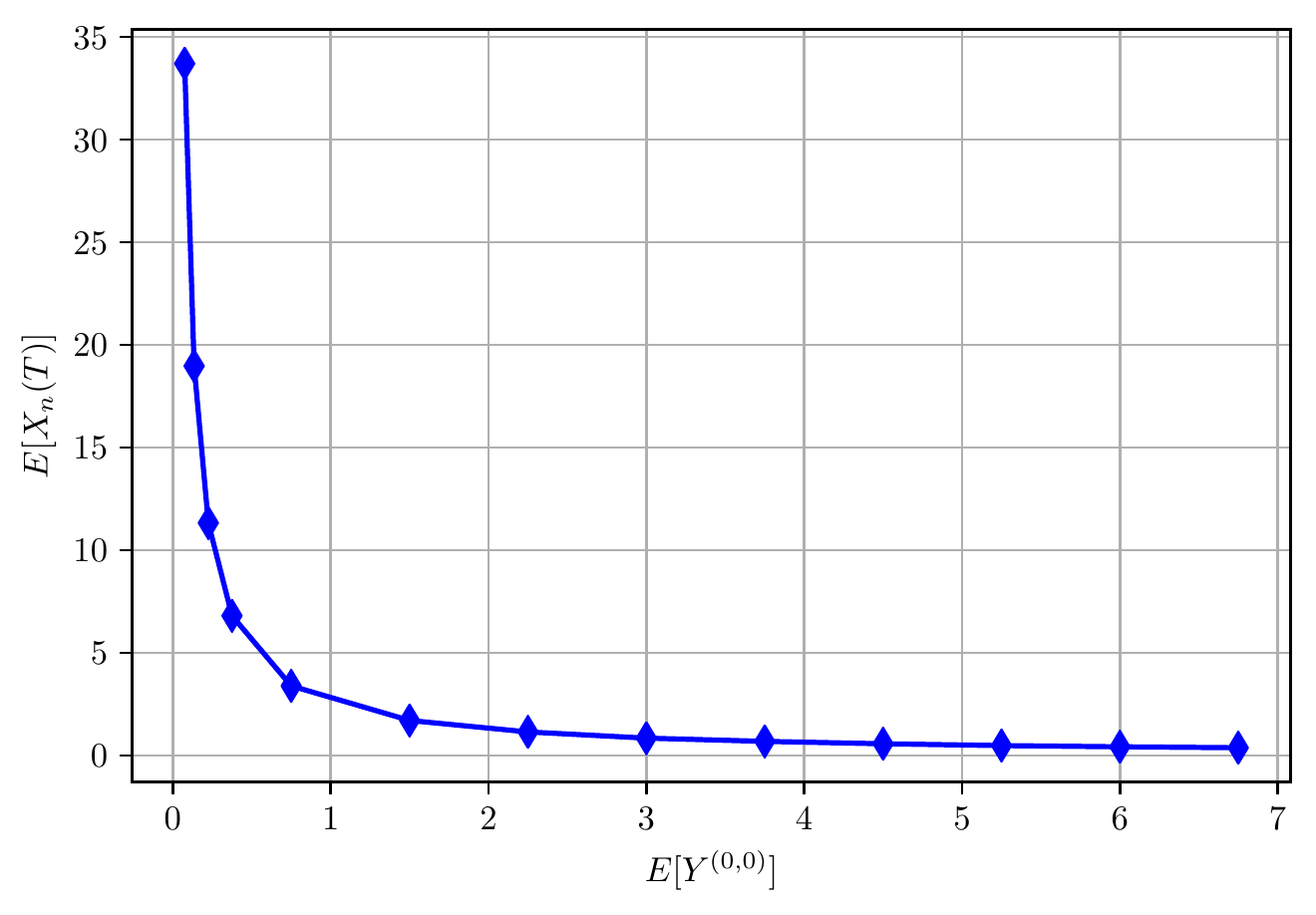}}
\caption{$\mathbb{E}[X_n(T)]$ in $3$-hop network with $Y^{(0,0)} \sim Pareto(3,m)$, $Y^{(0,1)} \sim Rayleigh(1)$, $Y^{(1,2)} \sim \chi^2(1)$ and $Y^{(2,3)} \sim Beta(2,3)$, as plotted against different values of $\mathbb{E}[Y^{(0,0)}]$ obtained by varying $m$.}
\label{fig:graph_age_vs_sourcemeanV}
\vspace*{-0.3cm}
\end{figure}

We first simulate the model in Fig.~\ref{fig:n_hop_modelV} for $n=3$, i.e., a $3$-hop model with links $(0,1),(1,2)$ and $(2,3)$ following the inter-renewal distributions: Rayleigh with scale $\sigma=1$, Chi-Square with degree of freedom $k=1$, and Beta with shape parameters $\alpha=2$, $\beta=3$, respectively. We update the source according to Pareto (Type I) distribution which has mean $\mathbb{E}[Y^{(0,0)}]=\frac{am}{a-1}$ for shape parameter $a$ and scale parameter $m$. We simulate the network for a large duration, $T=10^3$ and take average of $X_n(T)$ over $2\times10^5$ iterations to approximate $\mathbb{E}[X_n(T)]$ by the law of large numbers, which is used as a proxy for $\lim_{t \to \infty}\mathbb{E}[X_n(t)]$. Fig.~\ref{fig:graph_age_vs_sourcemeanV} shows the plot of $\mathbb{E}[X_n(T)]$ as a function of $\mathbb{E}[Y^{(0,0)}]$, obtained by varying scale parameter $m$ while keeping $a=3$ in Pareto distribution. The plot supports the theoretical prediction of $\frac{2.5479}{\mathbb{E}[Y^{(0,0)}]}$ from (\ref{eqn:nhop_limit_expec_age_node_n_V}).

\begin{figure}[t]
\centerline{\includegraphics[width=0.9\linewidth]{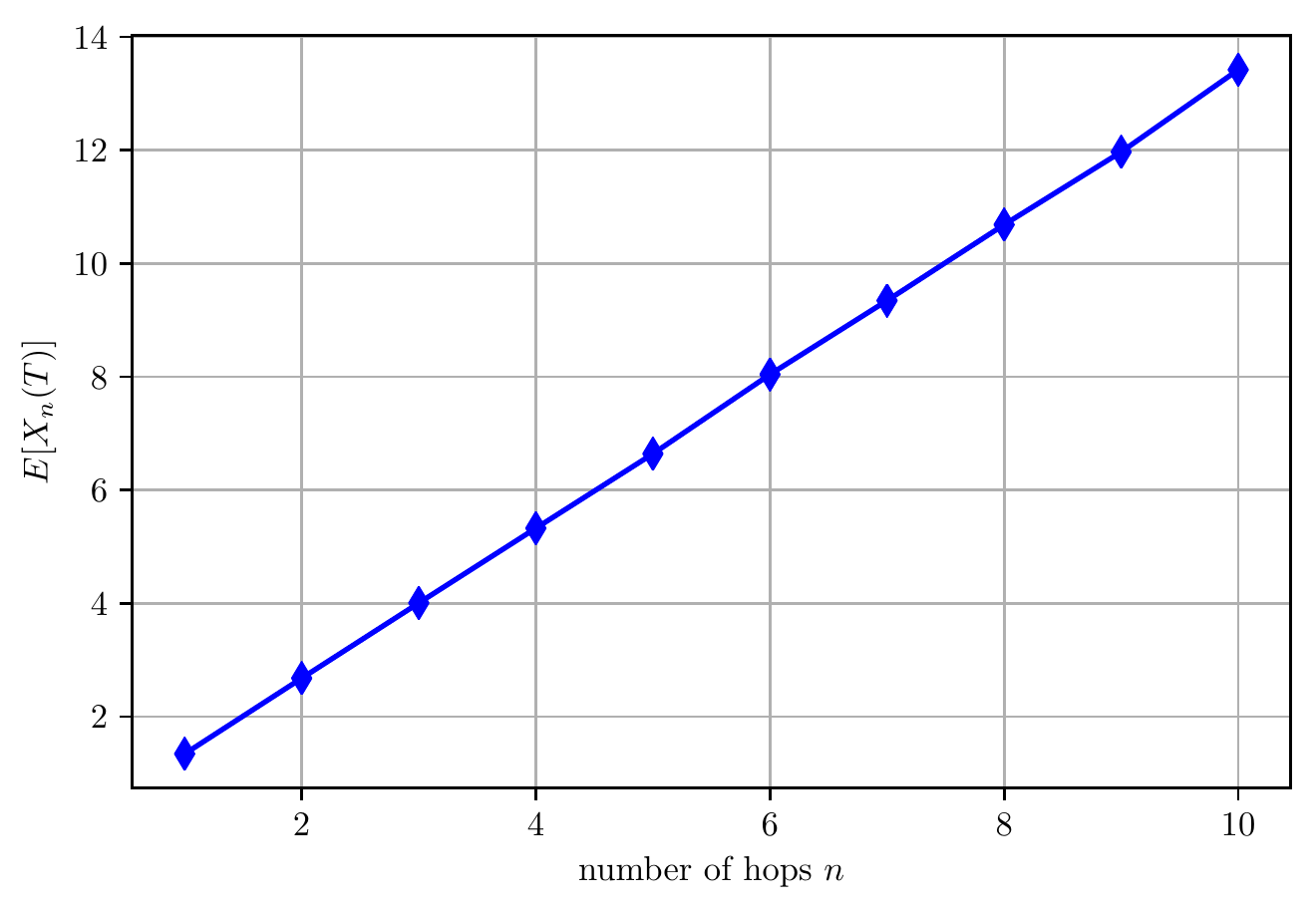}}
\caption{$\mathbb{E}[X_n(T)]$ in $n$-hop network with $Y^{(i,i+1)} \sim \mathcal{U}_{[0,2]}$ and $Y^{(0,0)} \sim Pareto(3,\frac{1}{3})$.}
\label{fig:graph_age_vs_no_hopsV}
\vspace*{-0.3cm}
\end{figure}

\begin{figure}[h!]
\centerline{\includegraphics[width=0.9\linewidth]{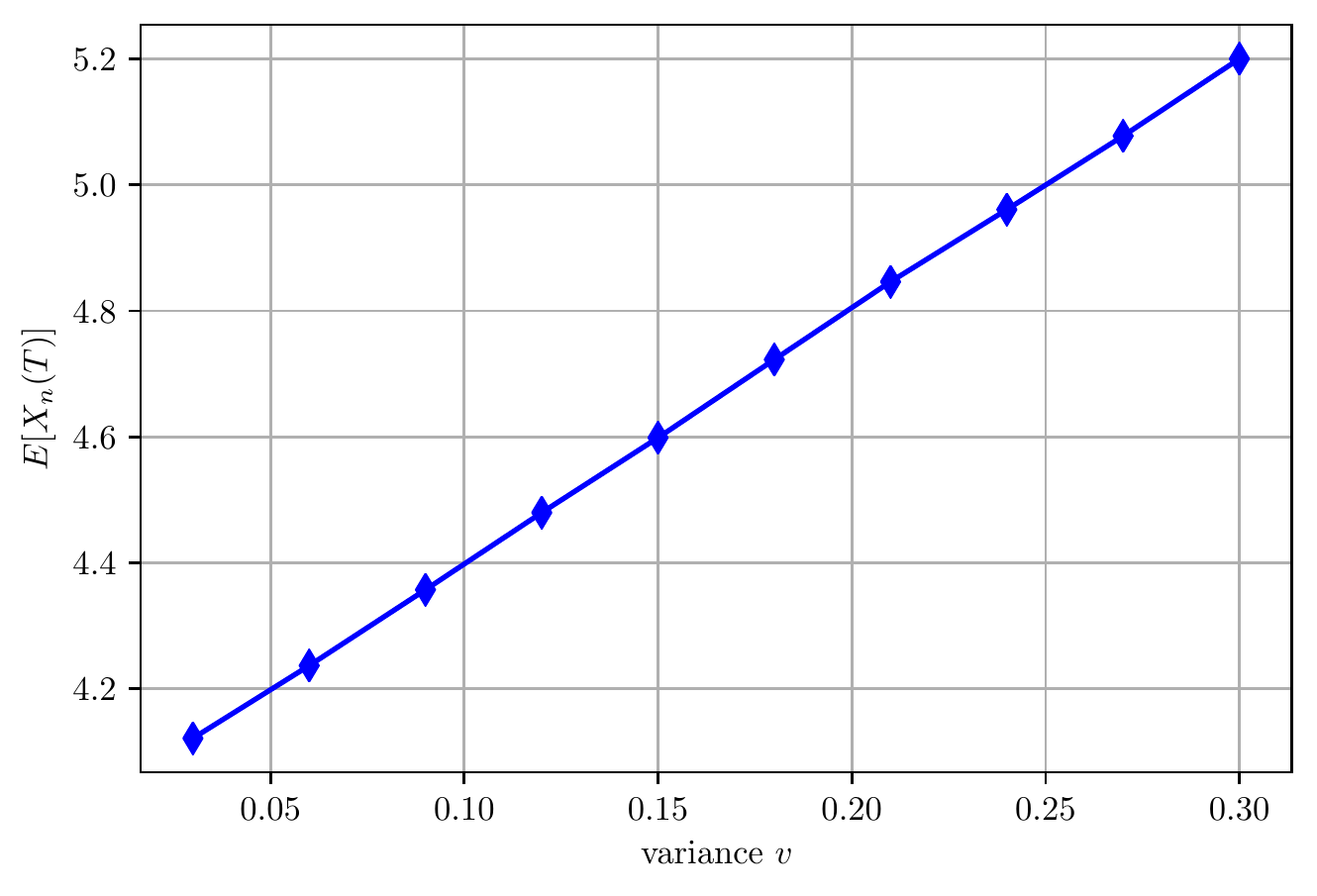}}
\caption{$\mathbb{E}[X_n(T)]$ in $4$-hop network with $Y^{(i,i+1)} \sim \mathcal{U}_{[1-\sqrt{3v},1+\sqrt{3v}]}$ and $Y^{(0,0)} \sim Pareto(3,\frac{1}{3})$.}
\label{fig:age_vs_varianceV}
\vspace*{-0.4cm}
\end{figure}

Next, we simulate an $n$-hop network where update intervals of all links $(i,i+1)$ follow uniform distribution on the interval $[0,2]$, i.e., $Y^{(i,i+1)} \sim Y \sim \mathcal{U}_{[0,2]}$, such that $\frac{\mathbb{E}\left[Y^2\right]}{2\mathbb{E}[Y]}=\frac{2}{3}$, and the source gets updated according to Pareto (Type I) distribution with $a=3$ and $m=\frac{1}{3}$, giving $\mathbb{E}[Y^{(0,0)}]=0.5$. We plot $\mathbb{E}[X_n(T)]$ as a function of $n$ in Fig.~\ref{fig:graph_age_vs_no_hopsV}. The linearity of the graph with the number of hops $n$ in Fig.~\ref{fig:graph_age_vs_no_hopsV} demonstrates the additive structure of the age at the end user as found in (\ref{eqn:nhop_limit_expec_age_node_n_V}). Since all links have the same distribution for inter-update times, the graph in Fig.~\ref{fig:graph_age_vs_no_hopsV} follows a linear equation in $n$ as $\lim_{t \to \infty}\mathbb{E}[X_n(t)]= \frac{4}{3}n$, as predicted by (\ref{eqn:nhop_limit_expec_age_node_n_V}).

Finally, we simulate a $4$-hop network, where inter-update times on all links $(i,i+1)$ follow the uniform distribution $Y^{(i,i+1)} \sim Y \sim \mathcal{U}_{[1-\sqrt{3v},1+\sqrt{3v}]}$, such that $\mathbb{E}\left[Y\right]=1$ and $Var\left[Y\right]=v$, with $Y^{(0,0)}\sim Pareto(3,\frac{1}{3})$, $\mathbb{E}[Y^{(0,0)}]=0.5$. Note that for fixed mean $1$, the maximum value of $v$ is $\frac{1}{3}$ to ensure that the probability distribution has non-negative support. Fig.~\ref{fig:age_vs_varianceV} shows that for fixed mean, $\mathbb{E}[X_n(T)]$ increases linearly with variance $v$, with $\lim_{t \to \infty}\mathbb{E}[X_n(t)]= \frac{1}{0.5}\times (4\times \frac{v+1}{2})=4v+4$, as predicted by (\ref{eqn:nhop_limit_expec_age_node_n_V}). 

\bibliographystyle{unsrt}
\bibliography{ref_priyanka}

\end{document}